\documentclass[10pt,journal,compsoc]{IEEEtran}
\usepackage[T1]{fontenc}
\usepackage[latin9]{inputenc}
\usepackage{amsmath,amsfonts,amssymb,amsthm}
\usepackage{algorithmic}
\usepackage[english]{babel}
\usepackage{graphicx}
\usepackage{times}
\usepackage[usenames,dvipsnames]{xcolor}
\usepackage{psfrag}
\usepackage[caption=false,font=footnotesize]{subfig}
\allowdisplaybreaks

\begin{document}

\title{Towards Low-Latency and Ultra-Reliable Virtual Reality}

\author{\IEEEauthorblockN{Mohammed~S.~Elbamby\IEEEauthorrefmark{1}, Cristina Perfecto\IEEEauthorrefmark{2},
Mehdi~Bennis\IEEEauthorrefmark{1}, and Klaus Doppler\IEEEauthorrefmark{4}\\
} \IEEEauthorblockA{\IEEEauthorrefmark{1}Centre for Wireless Communications, University
of Oulu, Finland. \\
 emails: \{mohammed.elbamby, mehdi.bennis\}@oulu.fi \\
}\IEEEauthorblockA{\IEEEauthorrefmark{2}University of the Basque Country (UPV/EHU),
Bilbao, Spain. \\
 email: cristina.perfecto@ehu.eus\\
} \IEEEauthorblockA{\IEEEauthorrefmark{4}Nokia Bell Labs, Sunnyvale, CA 94086, USA. \\
  email: klaus.doppler@nokia-bell-labs.com\\
 }}
\maketitle

\begin{abstract}
\noindent Virtual Reality (VR) is expected to be one of the killer-applications in 5G networks. However, many technical bottlenecks and challenges need to be overcome to facilitate its wide adoption. In particular, VR requirements in terms of high-throughput, low-latency and reliable communication call for innovative solutions and fundamental research cutting across several disciplines. In view of this, this article discusses the challenges and enablers for ultra-reliable and low-latency VR. Furthermore, in an interactive VR gaming arcade case study, we show that a smart network design that leverages the use of mmWave communication, edge computing and proactive caching can achieve the future vision of VR over wireless.
\end{abstract}

\section*{Introduction}

The last two years have witnessed an unprecedented interest both from academia and industry towards mobile/wireless virtual reality (VR), mixed reality (MR), and augmented reality (AR). The ability of VR to immerse the user creates the next generation of entertainment experiences, MR and AR promise enhanced user experiences and will allow end-users to raise their head from smartphone screens. 5G encompasses three service categories: enhanced mobile broadband (eMBB), massive machine-type communication (mMTC), and ultra-reliable and low-latency communication (URLLC). Mobile VR, MR and AR applications are very much use case specific and sit at the crossroads between eMBB and URLLC seeking multiple Gbps of data uniformly delivered to end-users subject to latency constraints. It is well known that low latency and high reliability are conflicting requirements \cite{Mogensen_5Gtradeoffs_2014}. Ultra-reliability implies allocating more resources to users to satisfy high transmission success rate requirements, which might increase latency for other users. Smart network designs are required to realize the vision of interconnected VR/AR, characterized by smooth and reliable service, minimal latency, and seamless support of different network deployments and application requirements. 

\subsection*{Wireless and Mobile VR, MR and AR}

In essence VR, MR and AR differ in the proportion in which digital content is mixed with reality. Both AR and MR incorporate some aspects of the real environment around the user: while real elements are the main focus for AR, virtual elements play a leading role in MR. To accomplish their goal, AR or MR glasses and wearables need not block out the world around, they will overlay digital layers to the current view of the user. The human eye is very sensitive to incorrect information. In order to ``feel real'', the AR or MR system needs to build a 3D model of the environment to place virtual objects in the right place and handle occlusions. In addition, the lighting of the object needs to be adjusted to the scene. Conversely, VR refers to a 100\%
virtual, simulated experience. VR headsets or head mounted displays (HMD) cover the user's field of view (FOV) and respond to eye tracking and head movements to shift what the screen displays accordingly. That is, in VR the only links to the outside real world are the various inputs arriving from the VR system to the senses of the user that are instrumental in adding credibility to the illusion of living inside the virtually replicated location.

The ultimate VR system implies breaking the barrier that separates both worlds by being unable to distinguish between a real and synthetic fictional world \cite{ejder_VR_2017}. An important step in this direction is to increase the resolution of the VR system to the resolution of the human eye and to free the user from any cable connection that limits mobility and that, when in touch with the body, disrupts the experience.

Up until now the use of untethered VR HMDs has been relegated to simple VR applications and discreet to low quality video streaming delivered through smartphone headsets such as Samsung Gear VR, or cost efficient ones such as the Google Cardboard. Meanwhile, HDMI connection through 19-wire cable has been favored for PC-based premium VR headsets such as Oculus Rift, HTC Vive or PlayStation VR. The reason can be found in the latency-sensitivity (latency of rendered image of more than 15 ms can cause motion sickness) and the resource \textendash communications and computing\textendash{} intensiveness nature of VR systems. In addition, even premium VR headsets still have only a limited resolution of 10 pixels per degree, compared to 60 pixels per degree with clear (20/20) visual acuity of the human eye. Hence, HD wireless/mobile VR is doubly constrained. It is computing constrained, as GPU power in HMDs is limited by the generated heat in powering these devices and by the bulkiness and weight of the headset itself. Second, it is constrained by the bandwidth limitations of current wireless technologies, operating below 6 GHz, and the resulting inability to stream high resolution video \textendash 8K and higher\textendash{} at high frame rate \textendash over 90 frames per second (fps)\textendash. The success of wireless VR hinges on bringing enough computing power to the HMD via dedicated ASICS or to the cloud or fog within a latency budget. Yet, recent developments from the VR hardware industry could deliver to the market first commercial level standalone VR headgears in 2018 even if still with limited resolution.

A manifold of technological challenges stemming from a variety of disciplines need to be addressed to achieve an interconnected VR experience. An interconnected VR service needs to handle the resource distribution, quality of experience (QoE) requirements, and the interaction between multiple users engaging in interactive VR services. It should also be able to handle different applications and traffic scenarios, for example, the aggregate traffic of an enterprise floor where participants share an MR workplace or an interactive gaming arcade, where each player is experiencing her own VR content. 

Therefore, in this paper we envision that the next steps towards the future interconnected VR will come from a flexible use of computing, caching and communication resources, a.k.a. the so called C$^{3}$ paradigm. To realize this vision, many trade-offs need to be studied. These range from the optimization of local versus remote computation, to single or multi-connectivity transmission while taking into account bandwidth, latency and reliability constraints.

\section*{Requirements and Big Challenges in wireless VR}\label{sec:Requirements}

From a wireless communication point of view, the extremely high data rate demands coupled with ultra-low latency and reliability are the main hurdles before bringing untethered VR into our everyday lives. In what follows, we will briefly introduce the bandwidth/capacity, latency and reliability requirements associated to several VR use cases. 

\subsection*{Capacity}

Current 5G or new radio (NR) system design efforts aim at supporting the upcoming exponential growth in data rate requirements from resource-hungry
applications. It is largely anticipated that a 1000-fold improvement in system capacity \textendash defined in terms of bits per second per square kilometer b/s/km$^{2}$\textendash{} will be needed. This will be facilitated through increased bandwidth, higher densification, and improved spectral efficiency. Focusing on VR technology, a back-of-the-envelope calculation reveals that with each of the human eyes being able to see up to 64 million pixels (150$^{\circ}$ horizontal and 120$^{\circ}$ vertical FOV, 60 pixels per degree) at a certain moment \cite{ejder_VR_2017}, and with 120~fps requirement to generate a real-like view, up to 15.5 billions of pixels per second are needed. By storing each colored pixel in 36 bits, and with the maximum of 1:600 video compression rate typically found in H.265 HEVC encoding, a required bit rate of up to 1~Gbps is needed to guarantee such quality.

The values above are clearly unrealizable in 4G. Actually, even early stage and entry-level VR, whose minimum data rate requirements are estimated to reach 100~Mbps\footnote{Corresponding to 1K and 2K VR resolution or equivalent 240 pixel lines and SD TV resolution respectively} will not be supported for multiple users in many deployments. Adding the required real time response for dynamic and interactive collaborative VR applications, it is not surprising that a significant ongoing research effort is geared towards reducing bandwidth needs in mobile/wireless VR, thereby shrinking the amount of data processed and transmitted. For example, in the context of 360$^{\circ}$ immersive VR video streaming, head movement prediction is used in \cite{qian_optimCell_2016} to spatially segment raw frames and deliver in HD only their visible portion. A similar approach is considered in \cite{ju_ultraWideVRStream_2017}, splitting the video into separated grid streams and serving grid streams corresponding to the FOV. Alternatively, eye gaze tracking is applied in \cite{doppler_EUCNC_2017} to deliver high resolution content only at the center of the human vision and to reduce the resolution and color depth in the peripheral field of view. Such a foveated 360$^{\circ}$ transmission has the potential to reduce the data rates to about 100~Mbps for a VR system with less than 10~ms round trip time including the rendering in the cloud. Yet, even if we allow only 5~ms latency for generating a foveated 360$^{\circ}$ transmission, existing networks cannot serve 100~Mbps to multiple users with reliable round trip times of less than 5~ms. Secondly, in today's networks computing resources are not available this close to the users. Therefore, there exists a gap between what current state of the art can do and what will be required as VR seeps into consumer space and pushes the envelop in terms of network requirements. In view of this, we anticipate that the millimeter wave (mmWave) communications will bridge the gap by facilitating the necessary capacity increase. 

\subsection*{Latency}

In VR environments, stringent latency requirements are of utmost importance for providing a pleasant immersive VR experience. The human eye needs
to perceive accurate and smooth movements with low motion-to-photon (MTP) latency, which is the lapse between a moment (e.g. head rotation) and a frame's pixels corresponding to the new FOV have been shown to the eyes. High MTP values send conflicting signals to the vestibulo-ocular reflex (VOR), a dissonance that might lead to motion sickness. There is broad consensus in setting the upper bound for MTP to less than 15-20~ms. Meanwhile, the loopback latency of 4G under ideal operation conditions is 25~ms. 

The challenge for bringing end-to-end latency down to acceptable levels starts by first understanding the various types of delays involved in such systems to calculate the joint computing and communication latency budget. Delay contributions to the end-to-end wireless/mobile VR latency include, sensor sampling delay, image processing or frame rendering computing delay, network delay (queuing delay and over-the-air delay) and display refresh delay. Sensor delay's contribution ($<$1 ms) is considered imperceptible by users, and display delay ($\approx$10-15~ms) is expected to drop to 5 ms \cite{mangiante_VREdge_2017}, which leaves 14 ms for computing and communication.

Both computing and communication delay serve as delay bottleneck in VR systems. Heavy image processing requires high computational power that is often not available in the local HMD GPUs. Offloading computing tasks to remote cloud servers significantly relieves the computing burden from the users' HMDs at the expense of incurring additional communication delay in both directions. Unlike MR and AR where uploading video streams to the cloud may be required, uplink communication delay due to offloading the computing task to the server is typically very small in VR, owing to the small amount of data needed, e.g., user tracking data and the interactive control decisions. However, the downlink delivery of the processed video frames in full resolution
can significantly contribute to the overall delay. Current online VR computing can take as much as 100 ms and communication delay (edge of network to server) reach 40 ms. Therefore, relying on remote cloud servers is a more suitable approach for low-resolution non-interactive VR applications, where the whole 360$^{\circ}$ content can be streamed and the constraints on real-time computing are relaxed. Interactive VR applications require real-time computing to ensure responsiveness. Therefore, it is necessary to shrink the distance between the end users and the computing servers to guarantee minimal latency. Fog computing \textendash also known as mobile edge computing (MEC)\textendash, where the computation resources are pushed to the network edge close to the end users, serves as an efficient and scalable approach to provide low latency computing to VR systems. MEC is expected to reduce the communication delay to less than 1~ms in metropolitan areas. Another interesting scenario for the use of MEC, for AR, is provided in \cite{osvaldo_AR_MEC_2017} where, besides latency reduction, energy-efficiency is considered. The MEC resource allocation exploits inherent collaborative properties of AR: a single user offloads shared information on an AR scene to the edge servers which transmit the resulting processed data to all
users at once via a shared downlink.

\subsection*{Reliability}\label{subsec:reliable}

VR/AR applications need to consistently meet the stringent latency and reliability constraints. Lag spikes and dropouts need to be kept to a minimum, or else users will feel detached. Immersive VR demands a perceptible image-quality degradation-free uniform experience. This mandates error-robustness guarantees in different layers, spanning from the video compression techniques in the coding level, to the video delivery schemes in the network level. In wireless environments where temporary outages are common due to impairments in signal to interference plus noise ratio (SINR), VR's non-elastic traffic behavior poses yet an additional difficulty. In this regard, an ultra-reliable VR service refers to the delivery of video frames on time with high success rate. Multi-connectivity (MC) has been developed for enhancing data rates and enabling a reliable transmission. MC bestows diversity to reduce the number of failed handovers, dropped connections, and radio-link failure (RLF). MC can either operate using the same or separate carriers frequencies. In intra-frequency MC, such as in single frequency networks (SFN), multiple sources using the same carrier frequency jointly transmit signals to a user. Contrarily, inter-frequency MC, which includes carrier aggregation (CA), dual connectivity (DC) and the use of different wireless standards, leverages either single or various sources that employ multiple carrier frequencies simultaneously for the same purpose. Enhancing reliability always comes at the price of using more resources and may result in additional delays, for example at the PHY layer the use of parity, redundancy, and re-transmission will increase the latency. Also, allocating multiple sources for a single user could potentially impact the experienced latency of the remaining users. {Another important reliability aspect in 5G is the ultra-high success rate of critical low-throughput packets. In particular, a maximum packet error rate (PER) of 10$^{-5}$ is specified in the 3GPP standard. This correlates with the VR/AR tracking message signaling that has to be delivered with ultra-high reliability to ensure smooth VR service.} 

\begin{figure*}[ht!]
	\centering \includegraphics[width=.8\linewidth]{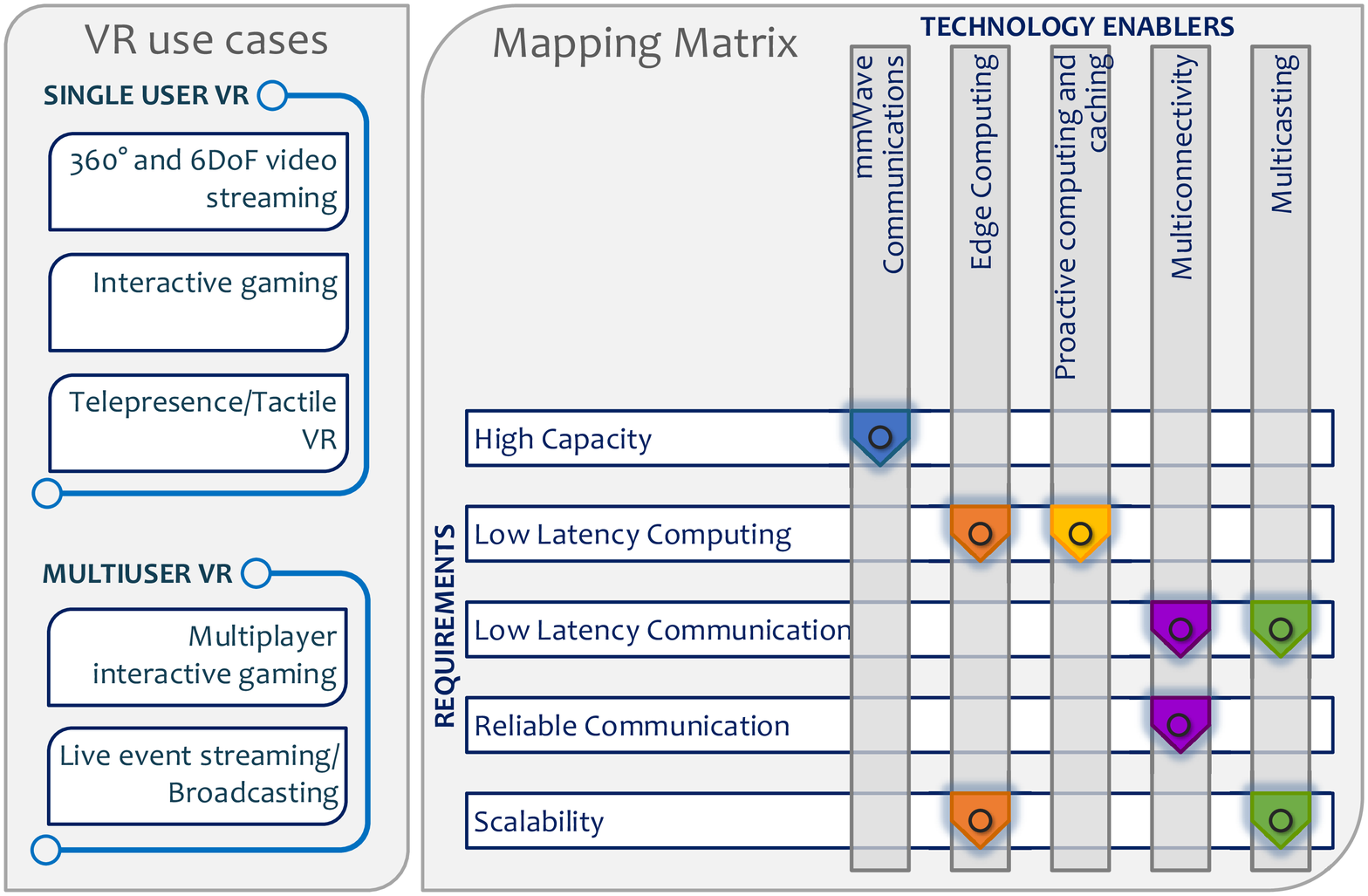}
	\caption{Single and multiple user VR use cases: requirements and enablers}
	\label{fig:use_cases} 
\end{figure*}

\section*{C$^{3}$: Enablers for URLLC in VR}

As outlined above, there is a substantial amount of work to be done to achieve a true immersive VR user experience. The VR QoE is highly dependent on stringent latency and reliability conditions. High MTP delays of 20 ms or more as well as distortions due to low data rate and resulting quality of the projected images, lead to motion sickness and affect the user visual experience. Hence, end-to-end delay and reliability guarantees are key elements of an immersive VR experience. Smart network designs that blend together and orchestrate communication, computing, and caching resources are sorely lacking. Figure~\ref{fig:use_cases} captures the foreseen requirements and the main technological enablers for both single and multiple user VR use cases. Next, we shed light on the envisioned roles of mmWave communications and MEC as two major thrusts of the future interconnected VR.

\subsection*{Millimeter Wave Communications}

mmWave communications is an umbrella term technically referring to any communication happening above 30 GHz. The possibilities offered by the abundance of available spectrum in these frequencies \textendash with channel bandwidths ranging from 0.85GHz at 28GHz band to up to 5 GHz at 73GHz\textendash{} are their main allure. In mmWaves directional communications need to be used to provide sufficient link budget \cite{mmwave_38}. mmWave propagation suffers from blockage as mmWaves do not propagate well through obstacles, including the human body which inflicts around 20-35 dB of attenuation loss, besides there are almost no diffractions. Best communication conditions are therefore met when there is a line-of-sight (LOS) path between the transmitter and the receiver with mainlobes of their antenna beams facing each other. However, partially blocked single reflection paths might still be usable at a reduced rate. Directionality and isolation from blockage significantly reduce the footprint of interference and make mmWave well-suited for dense deployments\footnote{Due to the broadness of the subject, we by refer interested readers in mmWave communications to the seminal work on mmWave for 5G \cite{rappaport_mmWWillWork_2013}, \cite{mmW_challenges_Rangan2015} on potentials and challenges of mmWave communications, and \cite{mmW_URLLC_Ford2017} on challenges for achieving URLLC in 5G mmWave cellular networks.}.

To find the transmitter and receiver beam combination or directional channel that maximizes the SINR, digital, hybrid or analog beamforming and beam-tracking techniques need to be applied. The beam training is able to track moving users in slowly time-variant environments and to circumvent blocked line of sight paths by finding strong reflectors. Especially in multiuser VR scenarios, the most likely source of sudden signal drop arises from either temporal blockages caused by user's own limbs (e.g. a raising hand) and bodies of surrounding players or from transmitter-receiver beam misalignment. In such cases, if the SINR drops below a certain threshold, an alternative directional channel discovering process needs to be triggered. However, beam-tracking through beam training for large antenna arrays involving big codebooks with narrow beams can incur large delays. For that reason, developing efficient beam training and beam-tracking techniques is an active area of research, specially for fast changing environments. For example, machine learning methods can be used to identify the most likely beam candidates to keep the disruption at a minimum. 

In this paper we advocate the use of MC to counteract the blockages and temporal disruptions of the mmWave channel. Specifically, non-coherent multisourced VR frame transmission will be showcased as a way to improve SINR and increase reliability of those links experiencing worse channel conditions. This approach is in line with the idea of overbooking radio and computing resources as a mean to protect against mmWave channel vulnerability \cite{barbarossa_overbookResources_2017}. The literature on the specific application of mmWave technologies for VR is scarce, with the exception of \cite{abari_cutCord_2016} in which for a single-user local VR scenario the use of a configurable mmWave reflector is proposed to overcome self-body blockage and avoid the need to deploy multiple transmitters. 

\subsection*{MEC Computing and Caching}

Rendering and processing VR HD video frames requires extensive computation resources. However, the need for compact and lightweight HMDs places a limit on their computational capabilities. Computation offloading is seen as a key enabler in providing the required computing and graphics rendering in VR environments. Users upload their tracking information, as well as any related data such as gaming actions or video streaming preferences to MEC servers with high computation capabilities. These servers perform the offloaded computing tasks and return the corresponding video frame in the downlink direction. 

Cloud computing servers are capable of handling CPU and GPU-hungry computing tasks due to their high computational capabilities.The distance to computing resources for real-time VR services is limited by the distance light travels during the maximum tolerable latency. The concept of edge computing strikes a balance between communication latency and computing latency by providing high computational resources close to the users. We envision edge computing as a key enabler for latency-critical VR computing services. However, to ensure efficient latency-aware computing services with minimal costs, server placement, server selection, computing resource allocation and task offloading decisions are needed. 

Indeed, providing stringent reliability and latency guarantees in real-time applications of VR is a daunting task. Dynamic applications, such as interactive gaming where real-time actions arrive at random, requires massive computational resources close to the users to be served on time. Therefore, the burden on real-time servers has to be decreased through facilitating proactive prefetching tasks and computing of the corresponding users' video frames. Recent studies have shown that VR gaming users' head movement can be predicted with high accuracy for upcoming hundreds of milliseconds \cite{qian_optimCell_2016}. Such prediction information can significantly help in relieving the burden on servers of real-time computing following users' tracking data. Based on estimated future pose of users, video frames can be proactively computed in remote cloud servers and cached in the network
edge or the users' HMDs, freeing more edge servers for real-time tasks. 

In addition to predicting user's movement, application-specific actions and corresponding decisions can be also proactively predicted. Since humans' actions are correlated, studying the popularity of different actions and their impact on the VR environment can facilitate in predicting the upcoming actions. Accordingly, subject to the available computing and storage resources, video frames that correspond to the speculated actions can be rendered and cached \cite{lee_outatime_2015}, ensuring reliable and real-time service. 

\begin{figure*}
\centering 
\includegraphics[width=.7\linewidth]{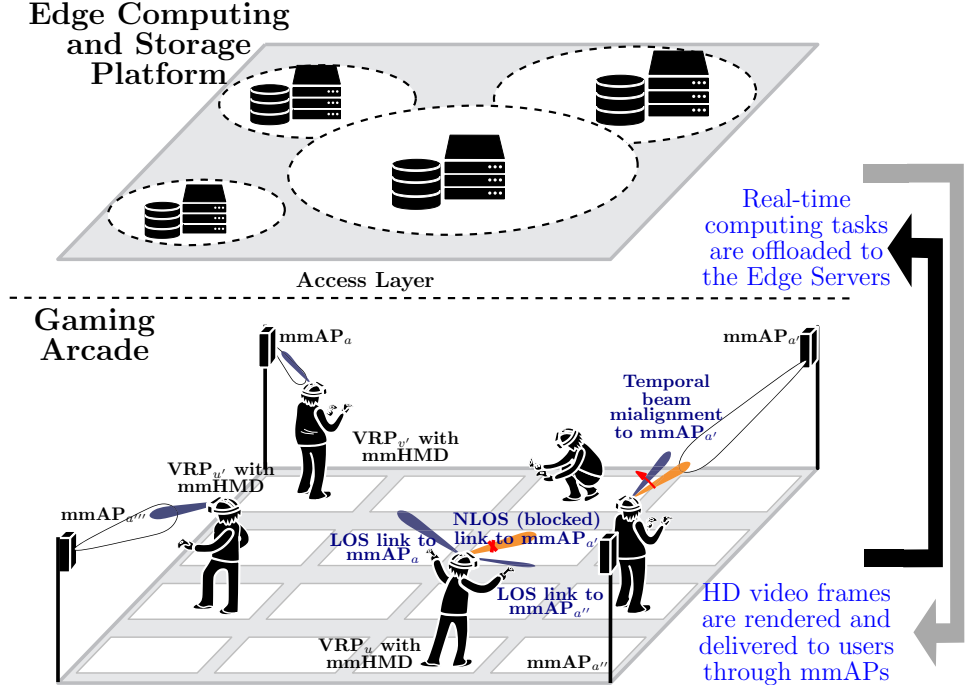}
\caption{Interactive VR gaming arcade with mmWave APs and edge computing network architecture.}
\label{fig:gaming_arcade}
\end{figure*}

\section*{Use Case: An Interactive VR Gaming Arcade}
\subsection*{Scenario Description}\label{subsec:scenario}

In this section, we investigate the use of C$^{3}$ to assess the URLLC performance of a multiplayer immersive VR gaming scenario. Such experience requires very low latency in order to synchronize the positions and interactions (input actions) of a group of players.

We consider an indoor VR gaming arcade where virtual reality players (VRPs) equipped with wireless mmWave head-mounted VR displays (mmHMD) are served by multiple mmWave band access points (mmAP) operating in 60-GHz indoor band\footnote{We remark that the cellular indoor 60 GHz scenario is one use case among many others. Our proposed approach to jointly combine edge computing with caching and mmWave communications leveraging multi-connectivity holds also for outdoor use and for any other mmWave band, e.g. for the 73 GHz licensed band, if wireless propagation particularities are appropriately addressed.}. VRPs move freely within the limits of individual VR pods, in which their movement in the physical space is tracked and mapped into the virtual space. Moreover, players' \emph{impulse actions} during the interactive gaming arrive at random, each of which is impacting the game play, and correspondingly the video frame content of a subset of the VRPs. 

mmAPs are connected to an edge computing network, consisting of multiple edge computing servers and a cache storage unit as illustrated in Figure \ref{fig:gaming_arcade}, where real-time tasks of generating users' HD frames can be offloaded based on the players' tracking data, consisting of their 6D pose and gaming impulse actions. In addition to real-time computing, we assume that the MEC network is able to predict users poses within a prediction window \cite{qian_optimCell_2016} to proactively compute and cache their upcoming video frames. A player can receive and display a proactively computed frame as long as no impulse action that impacts her arrives. The arrival intensity of impulse actions is assumed to follow a Zipf popularity distribution
with parameter $z$ \cite{ejder_VR_2017}. Accordingly, the arrival rate for the $i^{\textrm{th}}$ most popular action is proportional to $1/i^{z}$. The arrival of impulse action $i$ impacts the game play of a subset of players $\mathcal{U}_{i}$. The impact of the impulse actions on the VRPs' game play, namely, the \emph{impact matrix,} is defined as $\Theta=[\theta_{ui}]$, where $\theta_{ui}=1$ if $u\in\mathcal{U}_{i}$, and $\theta_{ui}=0$ otherwise\footnote{An example of an impulse action is a player firing a gun in a shooting game. As the game play of a subset of players is affected by this action, a video frame that has been already computed for any of them needs to be rendered again.}. A set of default parameters\footnote{We consider $4$ mmAPs, $4$ servers, $16$ players, $100$ impulse actions with popularity parameter $z=0.8$, and $10$ dBm mmAP transmit power.} are used for simulation purposes unless stated otherwise. 

When the game play starts, the MEC server keeps track of the arriving impulse actions and builds a popularity distribution of the action set. To keep up with the game dynamics, video frames that correspond to the most popular upcoming actions are computed and cached, subject to computing and storage constraints. 

\subsection*{Proposed Solution}

After the HD frames are rendered, the mmAPs schedule wireless DL resources to deliver the resulting video frames. As the delay of UL transmission to send the tracking data is typically small, we focus on the effect of computation delay in the edge servers and the DL communication delay. Scheduling is carried out such that the stringent latency and reliability constraints are met. In particular, the following probabilistic constraint on the frame delivery delay is imposed: 

\begin{equation}
\Pr(D_{\textrm{comm}}(t)+D_{\textrm{comp}}(t)\geq D_{\textrm{th}})\leq\epsilon,\label{eq:prob_const}
\end{equation}
which indicates that the probability that the summation of communication and computing delay at time instant $t$ exceeds a delay threshold value $D_{\textrm{th}}$ should be kept within a low predefined rate $\epsilon$. 

\begin{figure*}[ht!]
	\begin{minipage}[t]{0.47\linewidth}%
		\hspace{-1.1cm}
		\includegraphics[scale=0.76]{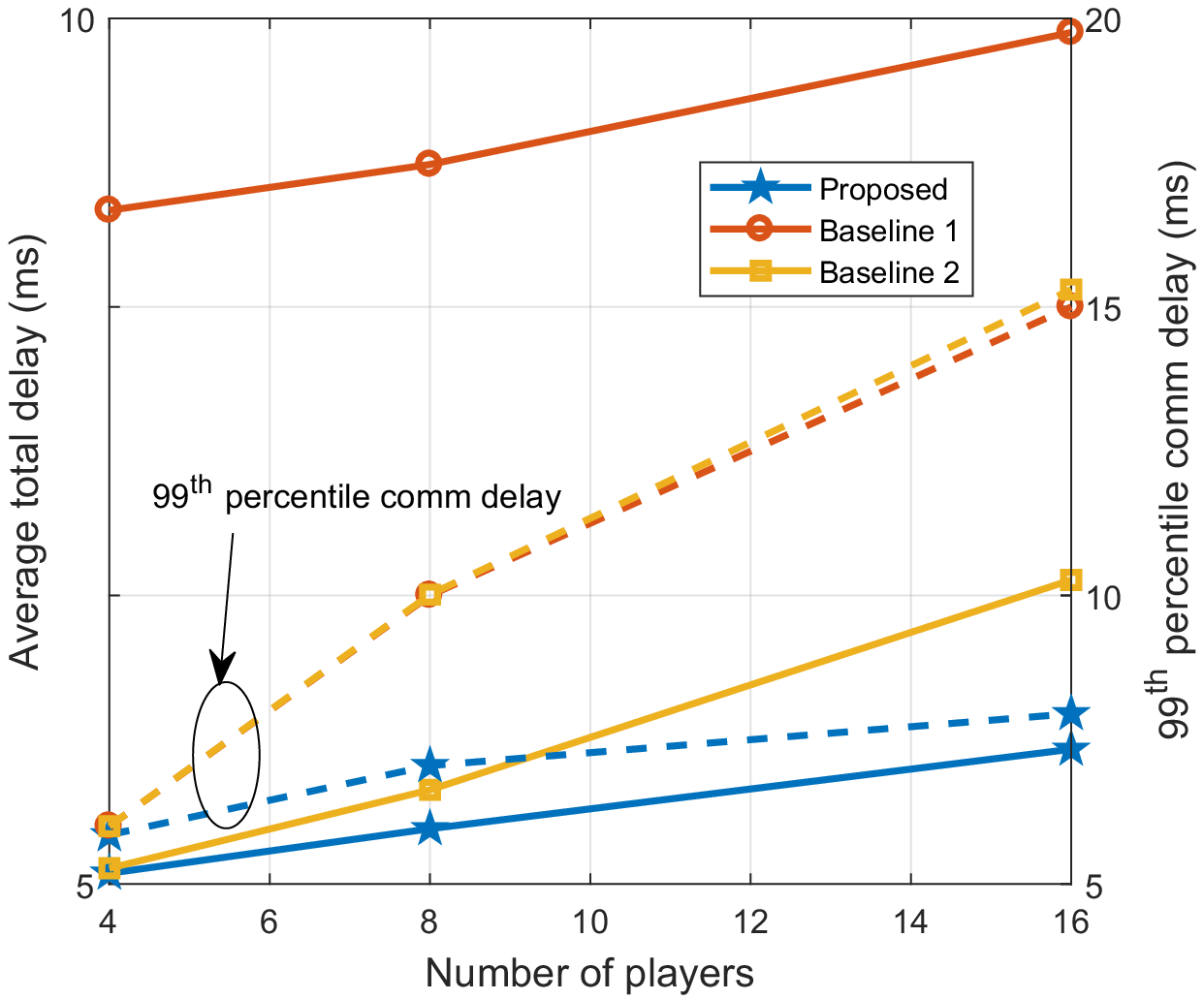}\caption{Average delay and $99^{\textrm{th}}$ percentile communication delay performance as the number of players varies, with 4 mmAPs and 8 MEC servers. The 99\% confidence level margin of error (ME) is $0.01088\leq \text{ME} \leq 0.0208$ ms for all configurations above.}
		\label{fig:fig3}
	\end{minipage}
	\begin{minipage}[t]{0.03\linewidth}%
		\hfill
	\end{minipage}
	\begin{minipage}[t]{0.47\linewidth}%
		\hspace{-.5cm}
		\includegraphics[scale=0.668]{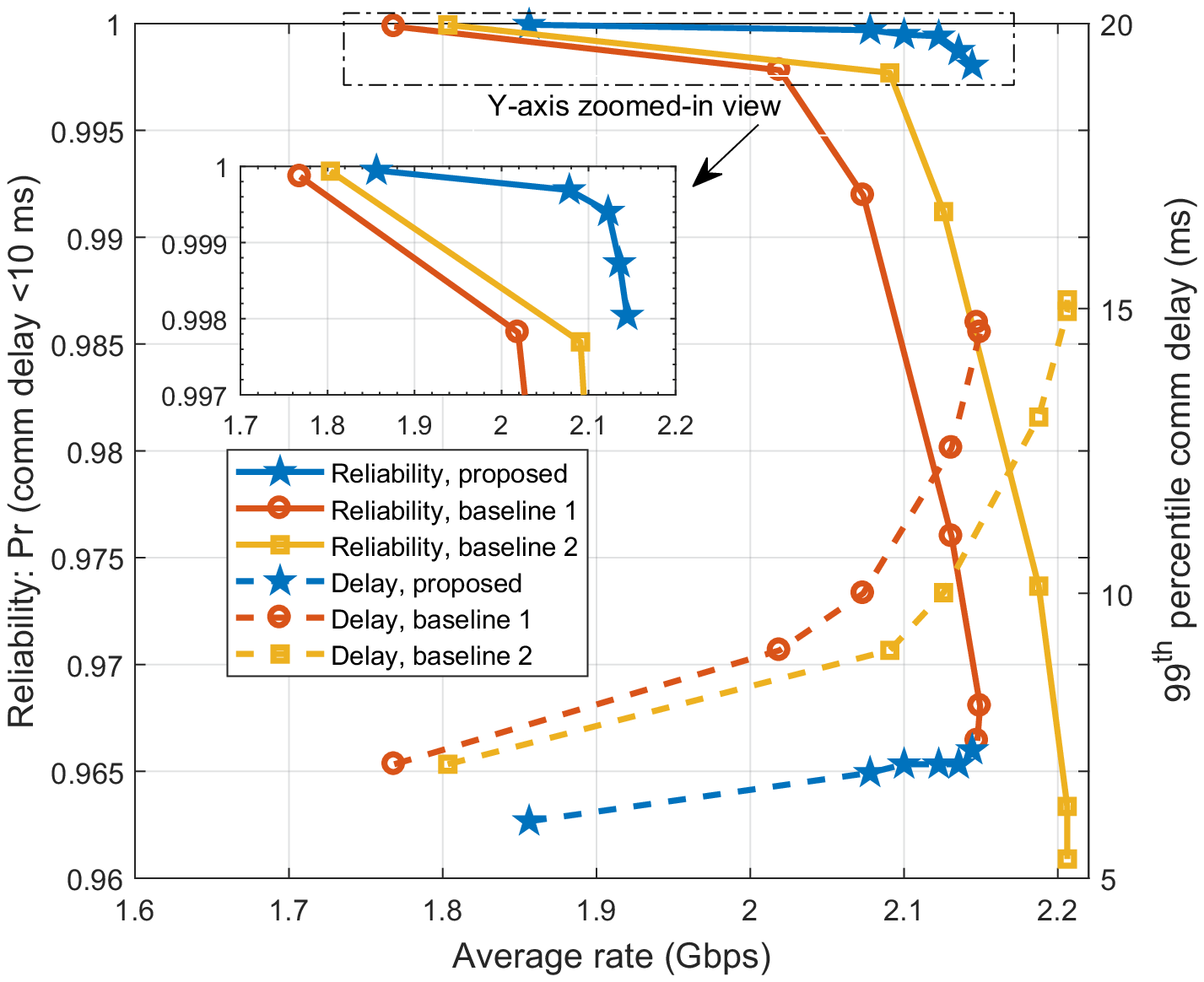}\caption{The tradeoffs between reliability performance (in terms of ratio of frames delivered within a given delay threshold), average service rate, and $99^{\textrm{th}}$ percentile communication delay, with 4 mmAPs, 8 MEC servers, and 16 VRPs.}\label{fig:fig4}
	\end{minipage}
\end{figure*}

To maintain a smooth game play in case of unsuccessful HD frame delivery, users perform local computing to generate a low resolution version of the required frame. In this regard, we propose an optimization framework to maximize the successful HD frame delivery subject to reliability and latency constraints. First, a joint proactive computing and caching scheme is developed to render users' HD frames in the network edge. HD frames that corresponds to users upcoming movement and head rotation and the estimated popular actions are proactively computed and cached. The proposed scheme schedules computing tasks following different priority levels, in which real-time computing is prioritized first in order to process current frames that are affected by randomly arriving game actions. Subsequently, subject to computing and storage resource constraints, the future HD frames are computed
and cached. 

Following the computation of HD frames, a matching algorithm based on the Deferred Acceptance (DA) matching \cite{gale_shapley} is considered
to allocate mmWave transmission resources to users. Matching preferences are selected such that reliability and latency constraints are met.
mmAPs preference over user requests are to achieve the latency constraint in (1), by prioritizing requests of users with tight latency deadlines.
User preferences over different mmAP aim to maximize user data rate, whereas dual-connectivity is considered by allowing users with an average rate below the rate threshold to be matched to a pair of mmAPs.

\begin{figure*}[h!]
	\centering
	\includegraphics[scale=0.65]{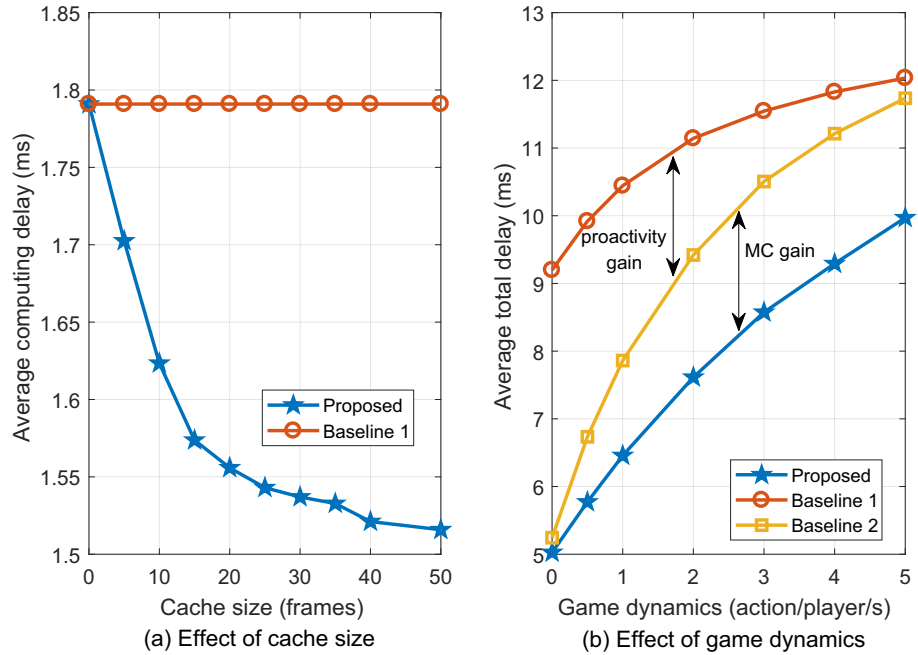}
	\caption{Average delay performance. (a) Average computing delay as the cache size increases, with 4 mmAPs, 8 MEC servers, and 8 VRPs; for all cache sizes and solution configurations above, ME for computing delay is $0.02005\leq \text{ME} \leq 0.02470$ ms. (b) Average total delay as the game traffic dynamics increase, with 4 mmAPs, 8 MEC servers, and 16 VRPs; for all considered game dynamics and solutions above, ME of the average total delay is	$0.00651\leq \text{ME} \leq 0.02469$ ms.}
	\label{fig:delay_results}
\end{figure*}

Next, we show and analyze the results of the proposed approach obtained from extensive system-level simulations. For the sake of comparison, we also plot two baseline schemes:  Baseline~1 with reactive computing (in which all computing is carried out in real-time) and Baseline~2 with proactive computing; neither Baseline 1 nor Baseline 2 have MC capability. The results therein have been averaged over 50 random game play instance topologies. Moreover, to give and idea of the size of the confidence intervals  99\% confidence level margin of errors (ME) have been computed and lowest and highest ME from all the possible configurations are provided.

\subsection*{Latency Performance}

First, we show the delay performance of the proposed approach with different number of players, each of which has a rate requirement of $2$ Gbps. By setting the parameters in (1) to $D^{\textrm{th}}=20$ ms and $\epsilon=0.01$ to reflect the motion sickness limit, we plot the average total delay as well as the $99^{\textrm{th}}$ delay performance of the proposed approach against the baseline schemes. From Figure~\ref{fig:fig3}, we can see that the proposed approach significantly minimizes the service delay in different network conditions. Moreover, by looking into the $99^{\textrm{th}}$ percentile communication delay, we find that the proposed scheme outperforms the proactive Baseline~2 scheme by leveraging MC to minimize the latency of wireless frame delivery. 

\subsection*{Reliability, Latency and Rate Tradeoffs}

Next, we show the tradeoffs of reliability, latency and service rate performance of the proposed scheme. Different results are obtained by varying the latency threshold in (1), while setting $\epsilon=0.01$ and the number of players to $16$. Reliability is measured by the probability of experiencing a communication delay below a threshold of $10$ ms. In Figure~\ref{fig:fig4}, we can see that there exists a tradeoff between the user data rate and the reliability and communication latency. Imposing stringent latency constraint guarantees achieving high reliability by serving requests with tight delay bounds. This comes at the expense of experiencing lower service rate and hence, lower frame quality.

\subsection*{Average Delay Performance}

Figure~\ref{fig:delay_results} compares the total delay performance of the proposed scheme against the reactive and proactive baseline schemes in different network conditions. In Figure~\ref{fig:delay_results}-a, it is shown that as the cache size increases, the average computing delay is significantly reduced. This reduction is due to caching more HD frames following popular game actions, which minimizes the computation delay as compared to the reactive baseline scheme. The effect of both proactivity and MC on the delay performance is also evident in Figure~\ref{fig:delay_results}-b, where the total VR service delay is plotted against the game dynamics, defined as the impulse action arrival intensity (action per player
per second). For all schemes, higher delay values are experienced as the game dynamics increase, due to having to process more frames in real-time. The proposed scheme is shown leverage both proactivity and MC to minimize the service delay in different gaming traffic conditions.

\section*{Conclusion}

In this article, we have discussed the main requirements for an interconnected wireless VR, MR and AR. We have highlighted the limitations of today's
VR applications and presented the key enablers to achieve the vision of future ultra-reliable and low latency VR. Among these enablers, the use of mmWave communication, mobile edge computing and proactive caching are instrumental in enabling this vision. In this respect, our case study demonstrated the performance gains and the underlying tradeoffs inherent to wireless VR networks. 

\section*{Acknowledgments}\label{sec:ack}
This research was partially supported by the Academy of Finland project CARMA, the NOKIA donation project FOGGY, the Thule Institute strategic project SAFARI and by the Spanish Ministerio de Economia y Competitividad (MINECO) under grant TEC2016-80090-C2-2-R (5RANVIR).


\section*{Biographies}
\enlargethispage{-13cm}
\vspace{-2.7cm}
\begin{IEEEbiographynophoto}{MOHAMMED S. ELBAMBY} (mohammed.elbamby@oulu.fi) received the B.Sc. degree (Hons.) in Electronics and Communications Engineering from the Institute of Aviation Engineering and Technology, Egypt, in 2010, and the M.Sc. degree in Communications Engineering from Cairo University, Egypt, in 2013. He is currently pursuing the Dr.Tech. degree with the University of Oulu. After receiving the M.Sc. degree, he joined the Centre for Wireless Communications, University of Oulu. His research interests include resource optimization, uplink and downlink configuration, fog networking, and caching in wireless cellular networks. He received the Best Student Paper Award from the European Conference on Networks and Communications in 2017.
\end{IEEEbiographynophoto}
\vspace{-2.5cm}
\begin{IEEEbiographynophoto}{CRISTINA PERFECTO} (cristina.perfecto@ehu.eus) is a Ph.D. student at the University of the Basque Country (UPV/EHU), Bilbao, Spain. She received her B.Sc. and M.Sc. in Telecommunication Engineering from UPV/EHU in 2000 where she is currently a college associate professor at the Department of Communications Engineering. Her research interests lie on millimeter wave communications and in the application of machine learning in 5G networks. She is currently working towards her Ph.D. focused on the application of multidisciplinary computational intelligence techniques in radio resource management for 5G.
\end{IEEEbiographynophoto}
\vspace{-2.5cm}
\begin{IEEEbiographynophoto}{MEHDI BENNIS} [S'07-AM'08-SM'15] (mehdi.bennis@oulu.fi) received his M.Sc. degree in Electrical Engineering jointly from the EPFL, Switzerland and the Eurecom Institute, France in 2002. \linebreak \newpage \noindent From 2002 to 2004, he worked as a research engineer at IMRA-EUROPE investigating adaptive equalization algorithms for mobile digital TV. In 2004, he joined the Centre for Wireless Communications (CWC) at the University of Oulu, Finland as a research scientist. In 2008, he was a visiting researcher at the Alcatel-Lucent chair on flexible radio, SUPELEC. He obtained his Ph.D. in December 2009 on spectrum sharing for future mobile cellular systems. Currently Dr. Bennis is an Associate Professor at the University of Oulu and Academy of Finland research fellow. His main research interests are in radio resource management, heterogeneous networks, game theory and machine learning in 5G networks and beyond. He has co-authored one book and published more than 100 research papers in international conferences, journals and book chapters. He was the recipient of the prestigious 2015 Fred W. Ellersick Prize from the IEEE Communications Society, the 2016 Best Tutorial Prize from the IEEE Communications Society and the 2017 EURASIP Best paper Award for the Journal of Wireless Communications and Networks. Dr. Bennis serves as an editor for the IEEE Transactions on Wireless Communication.
\end{IEEEbiographynophoto}
\begin{IEEEbiographynophoto}{KLAUS DOPPLER} (klaus.doppler@nokia-bell-labs.com) is heading the Connectivity Lab in Nokia Bell Labs and his research focus is on indoor networks. In the past, he has been responsible for the wireless research and standardization in Nokia Technologies, incubated a new business line and pioneered research on Device-to-Device Communications underlaying LTE networks. He received his PhD. from Aalto University School of Science and Technology, Helsinki, Finland in 2010 and his MSc. from Graz University of Technology, Austria in 2003. 
\end{IEEEbiographynophoto}
\vfill

\end{document}